\documentclass[a4paper,12pt]{spieman}  % use this instead for A4 paper
\usepackage{amsmath,amsfonts,amssymb}
\usepackage{graphicx}
\usepackage{setspace}
\usepackage{tocloft}
\usepackage{times}
\usepackage{bm} % for math
\usepackage{amssymb} % for math
\usepackage{amsmath} % for math
\usepackage{braket} %for brakets
\usepackage{mathrsfs} %for script letters 

\title{Characterizing the Nash equilibria of three-player Bayesian quantum games}

\author[a]{Neal Solmeyer}
\author[b]{Radhakrishnan Balu}
\affil[a]{Sensors and Electron Devices Directorate, Army Research Laboratory, Adelphi, MD, 21005-5069, USA}
\affil[b]{Computer and Information Sciences Directorate, Army Research Laboratory, Adelphi, MD, 21005-5069, USA}

\cftpagenumbersoff{figure}
\cftpagenumbersoff{table} 

\begin{document} 
\maketitle

\begin{abstract}
Quantum games with incomplete information can be studied within a Bayesian framework. We analyze games quantized within the EWL framework [Eisert, Wilkens, and Lewenstein, Phys Rev. Lett. 83, 3077 (1999)]. We solve for the Nash equilibria of a variety of two-player quantum games and compare the results to the solutions of the corresponding classical games. We then analyze Bayesian games where there is uncertainty about the player types in two-player conflicting interest games.  The solutions to the Bayesian games are found to have a phase diagram-like structure where different equilibria exist in different parameter regions, depending both on the amount of uncertainty and the degree of entanglement. We find that in games where a Pareto-optimal solution is not a Nash equilibrium, it is possible for the quantized game to have an advantage over the classical version. In addition, we analyze the behavior of the solutions as the strategy choices approach an unrestricted operation. We find that some games have a continuum of solutions, bounded by the solutions of a simpler restricted game. A deeper understanding of Bayesian quantum game theory could lead to novel quantum applications in a multi-agent setting. 
\end{abstract}

% Include a list of up to six keywords after the abstract
\keywords{Quantum games, Bayesian games}

% Include email contact information for corresponding author
{\noindent \footnotesize\textbf{*}Neal Solmeyer  \linkable{neal.e.solmeyer.civ@mail.mil} }

\begin{spacing}{1}   % use double spacing for rest of manuscript

\section{Introduction}
\label{sect:intro}  % \label{} allows reference to this section

Complex decision making tasks over a distributed quantum network, a network including entangled nodes, can be analyzed with a quantum game theory approach. Quantum games extend the applicability of classical games to quantum networks, which may soon be a reality. Quantum game theory imports the ideas from quantum mechanics such as entanglement and superposition, into game theory. The inclusion of entanglement leads to player outcomes that are correlated so that entanglement often behaves like mediated communication between players in a classical game. This can lead to a game that has different Nash equilibria with greater payoffs than the classical counterpart. The analysis of quantum games with entanglement can resemble the correlated equilibria of classical games. 

The entanglement is imposed by a referee, and acts like a contract that cannot be broken between the players, and can persist non-locally after the initial entanglement has been performed and communication forbidden. This is in contrast to classical correlated equilibria that rely on communication between the players, whose contracts can be broken, and cannot exhibit the non-local behavior associated with quantum mechanics. The correlations produced by entanglement can achieve probability distributions over the payoffs that are not possible in the classical game, even when mixed strategies are used.

 When interacting with a network, the agents will often have incomplete information about the other nodes. Quantum games with incomplete information can be treated within a Bayesian approach. With this approach in mind, we are interested in quantized games with classical priors, i.e. a statistical mixture of two quantum games. Detailed analysis of Bayesian quantum games can potentially lead to applications in quantum security protocols\cite{Maitra2015}, the development of distributed quantum computing algorithms\cite{Li2009}, or improving the efficiency of classical network algorithms \cite{Zabaleta2017}. Experiments have begun to demonstrate the results of quantum game theory in nuclear magnetic resonance \cite{Du2002}, quantum circuits in optical \cite{Zeilinger}, and ion-trap platforms \cite{Shuichi}, which, in some cases, i.e. optical, can be easily imagined on a distributed quantum network.

\subsection{Quantizing a classical game}

\begin{figure*}
\begin{center}
\begin{tabular}{c}
\includegraphics[width=1\columnwidth]{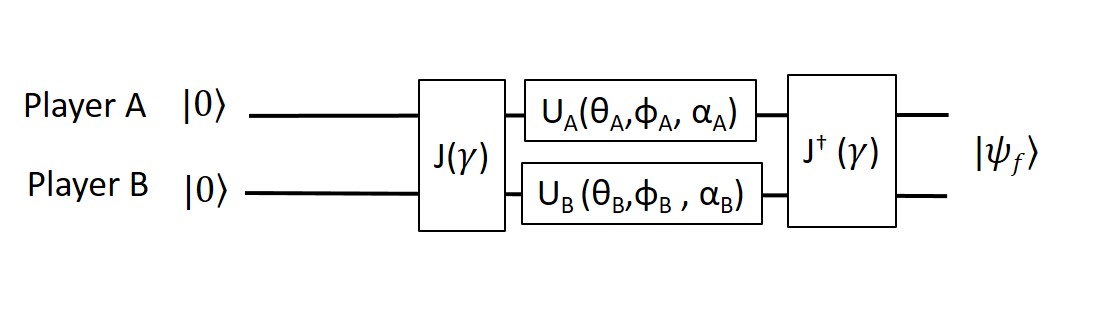}
\end{tabular}
\end{center}
\caption{\label{fig:QPD}Quantum circuit for two-player game }
\end{figure*}

To quantize a classical game, we follow the approach given in the seminal Einstein-Wilkens-Lewenstein scheme. The scheme goes as follows; both players qubits are initialized to the $\ket{0}$ state, an entangling operation, $J$, is applied, the players apply their strategy choice, $U_{A,B}(\theta,\phi,\alpha)$, an un-entangling operation is applied, the payoffs are determined from the probability distribution of the final state $\ket{\psi_f}$. This procedure can be encoded in the quantum circuit show in Figure \ref{fig:QPD}.

The amount of entanglement that occurs can be varied by varying the parameter $\gamma$ in the entangling operation: 

\begin{equation}
\hat{J}(\gamma)=e^{ i \gamma \hat{\sigma}_x \otimes \hat{\sigma}_x }= 
\begin{pmatrix}
Cos(\gamma/2) & 0 &0 &\imath \: Sin(\gamma/2)\\
0 & Cos(\gamma/2) &-\imath \: Sin(\gamma/2) &0) \\
0 & -\imath \: Sin(\gamma/2) &Cos(\gamma/2) & 0 \\
\imath \:Sin(\gamma/2) & 0 &0 & Cos(\gamma/2)
\end{pmatrix}
\label{eq:J} 
\end{equation}

At maximal entanglement,$\gamma=\pi/2$, this operation produces a Bell state, and at $\gamma=0$ is the identity operator.
The game is defined by setting the possible strategies of the players. For this we parametrize a single qubit rotation, $U$, with three parameters,$ (\theta,\phi,\alpha)$ in:

\begin{equation}
\hat{U}(\theta, \phi,\alpha)=
\begin{pmatrix}
e^{-\imath \phi} Cos(\theta/2) & e ^{\imath \alpha} Sin(\theta/2) \\
- e ^{-\imath \alpha} Sin(\theta/2)&e^{\imath \phi} Cos(\theta/2)
\end{pmatrix}
\label{eq:strat}
\end{equation} 

where $\theta \in [0,\pi],\phi \in [0,2\pi],\alpha \in [0,2\pi]$.

The outcome of the game is given by:

\begin{equation}
\ket{\psi_f(A,B)}= \hat{J}^{\dagger}(U_A \otimes U_B)\hat{J}\ket{00}
\label{eq:psif}
\end{equation}

And the average payoff $\langle\$\rangle$ is derived from the expectation values of a measurement performed at the end and the payoff vector $\$_j$
\begin{equation}
\langle\$^A(A,B) \rangle = \sum\limits_{j}\Braket{\psi_f(A,B) | \psi_f(A,B)}_j \$^A_j
\label{eq:payoff}
\end{equation}

There are four possible outcomes, $\{\ket{00},\ket{ 01}, \ket{10}, \ket{ 11}\}$. Correspondence to the classical game is made by associating each outcome as one of the classical strategy choices, such that $\ket{0}$ corresponds to Confess (C), and $\ket{1}$ corresponds to Defect (D), as is illustrated in the canonical prisoner’s dilemma game with payoff matrix shown in Table \ref{tab:PDmatrix}.

\begin{table}[ht]
\caption{Payoff matrix for prisoner's dilemma} 
\label{tab:PDmatrix}
\begin{center}       
\begin{tabular}{|c|c|c|} %% this creates two columns
%% |l|l| to left justify each column entry
%% |c|c| to center each column entry
%% use of \rule[]{}{} below opens up each row
\hline
\rule[-1ex]{0pt}{3.5ex}  $A| B_1$ & $ \Ket{0}(C)$ &$\Ket{1}(D)$\\
\hline\hline
\rule[-1ex]{0pt}{3.5ex} $ \Ket{0}(C)$  & $ (3,3)$ & $(0,5)$ \\
\hline
\rule[-1ex]{0pt}{3.5ex}  $\Ket{1}(D)$& $ (5,0)$ & $(1,1)$ \\
\hline
\end{tabular}
\end{center}
\end{table} 

\subsection{Forming a Bayesian game}

\begin{figure}
\includegraphics[width=1\columnwidth]{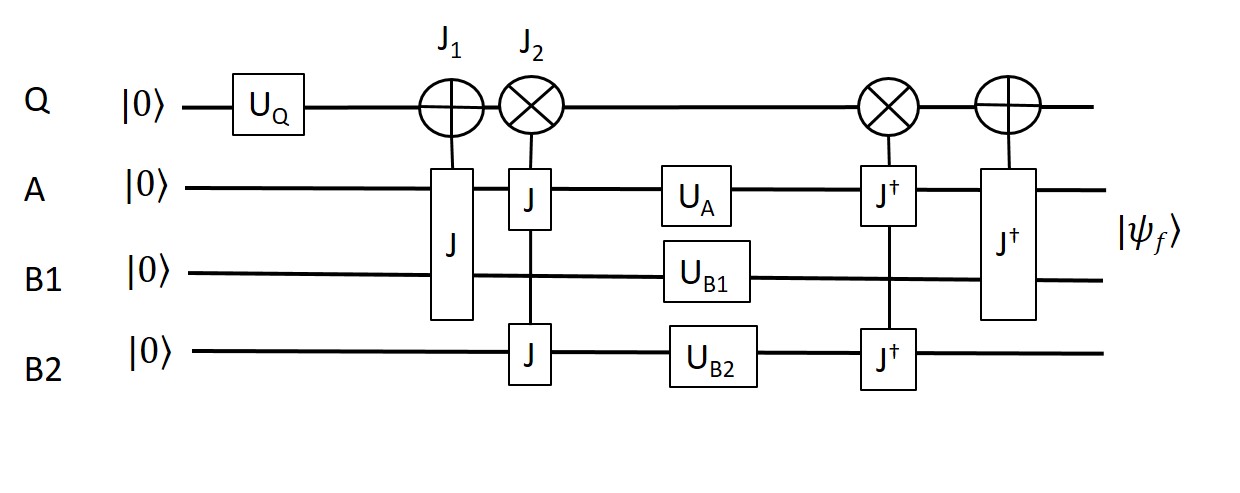}
\caption{\label{fig:quantumcircuit} Quantum circuit for Bayesian game}
\end{figure}

The Bayesian game is constructed with the protocol laid out by Harsanyi\cite{Harsanyi}. In the Bayesian game the players have incomplete knowledge about their opponent’s payoff matrices. This is represented by having the players receive a statistical mixture of different payoff matrices. Below we analyze games that are represented by two different payoff matrices. If, for example, player A’s payoff is the same in both matrices while player B’s vary, this represents player A having incomplete knowledge about player B’s preferences. If both have different payoffs, this could be interpreted as two players having incomplete knowledge about what game their playing. This game can be represented by the quantum circuit shown in Figure \ref{fig:quantumcircuit}.

$U_Q$ is a unitary operation on the control qubit. $J_1$ and $J_2$ are controlled entangling operations acting on $A$ and $B_1$ or$ A$ and $B_2$, depending on the state of the control qubit. This representation is equivalent to playing a statistical mixture of two quantum circuits shown in Figure \ref{fig:QPD} with different two-player games.

The average payoff for player A in the Bayesian game is given by:

\begin{equation}
\langle \$_A\rangle = \langle\$^A(A,B_1) \rangle (p)+ \langle\$^A(A,B_2) \rangle(1-p) 
\label{eq:Bpayoff}
\end{equation}

The B player's average payoff is still calculated according to Equation \ref{eq:payoff}.
\subsection{Solution Concepts}

The primary solution concept used in game theory is the Nash equilibrium. A Nash equilibrium is a set of strategies where neither player could benefit by unilaterally deviating. The payoff to the player’s at the Nash equilibrium represents a stable payoff in a repeated game or large ensemble, because it is self-enforcing.

There are refinements to the concept of a Nash equilibrium that are used to capture different types of games. Relevant to quantum games is the concept of a correlated equilibrium. A correlated equilibrium is a game where the player’s strategy choices are correlated in some way, such as reacting to advice or a contract, such that probability distributions are possible that are not in the image of the classical game with mixed strategies.  Entanglement acts to correlate the player’s outcomes in a similar way, except in quantum games the entanglement is imposed by a referee, and once the entanglement is produced, the player’s cannot break the contract. 

\subsection{Solution methods}
The method of best responses is used to find the Nash equilibria of the game. There have been analytic solutions found for certain cases of quantum games(cite), but with the aim to examine a wide range of games, including a Bayesian framework and asymmetric payoffs, and for these cases, analytic solutions remain elusive. Therefore we adopt a numerical procedure. 

First, the possible strategies must be chosen. Equation \ref{eq:strat} represents a completely arbitrary strategy choice. It is instructive to analyze the game with a more discretized set of strategies. We chose a step size for each parameter $(\Delta \theta, \Delta \phi, \Delta \alpha)$. Then we compile a list that contains all possible combinations of integer multiples of these steps, within the bounds of the parameters: 
\begin{equation}
V = \{U(0,0,\Delta \alpha), U(0,0,2\Delta \alpha), U(0,0,3\Delta \alpha)\ldots(0, \Delta \phi,0), U(0, \Delta \phi, \Delta \alpha)\ldots U(\pi,2\pi,2\pi)\} 
\label{eq:space}
\end{equation}
Where $V_i$ represents the ith element of the list. This set defines the possible strategies of a game. Several of these matrices are redundant, because for example, when $\theta = 0$, $\alpha$ is undefined. 

To construct the best response list for player A, for each possible strategy choice of player B, we compute the payoff for player A for each of their possible strategy choices. Then we select the elements which have the highest payoff, or best response, $U^*$. This produces a list of A’s best responses to each of B’s strategy choices: $\Omega_A = \{U^*_{Aj}, U_{Bj}\}$, where j ranges over all possible strategy choices. Then similarly B’s best response list is composed,$\Omega_B = \{U_{Aj}, U^*_{Bj}\}$. The Nash equilibria are given by the intersection of the best response functions: $\Omega_A \cap \Omega_B$. For the Bayesian game, this procedure is straightforwardly extended to three players. 

Many of the interesting features of the games we examine exist with the stepping parameters $(\pi, \pi/2, \pi/2)$, which has a total of 8 unique strategy choices. The majority of the data are presented with these stepping parameters. Next we examine the behavior of one game in more detail as we step finer in each of the strategy parameters. This eventually becomes computationally impractical as the step sizes get too small. For example, the stepping parameters $(\pi/8, \pi/8, \pi/8)$ yield 1824 unique strategy choices. The code in Mathematica with this many strategy choices takes ~1 hour to compute all of the Nash equilibria for the two-player game for all values of entanglement, making solutions to the Bayesian game impractical to find with the current method. 

\section{Results}
\subsection{Two-player game results}
Here we present the solutions for several two-player games found in the literature \cite{flitney, Avishai2012} and textbook games.  For the sake of space, we will not discuss the real world interpretration of these games, rather we focus on their mathematical propeties. 

\begin{figure}
\includegraphics[width=1\columnwidth]{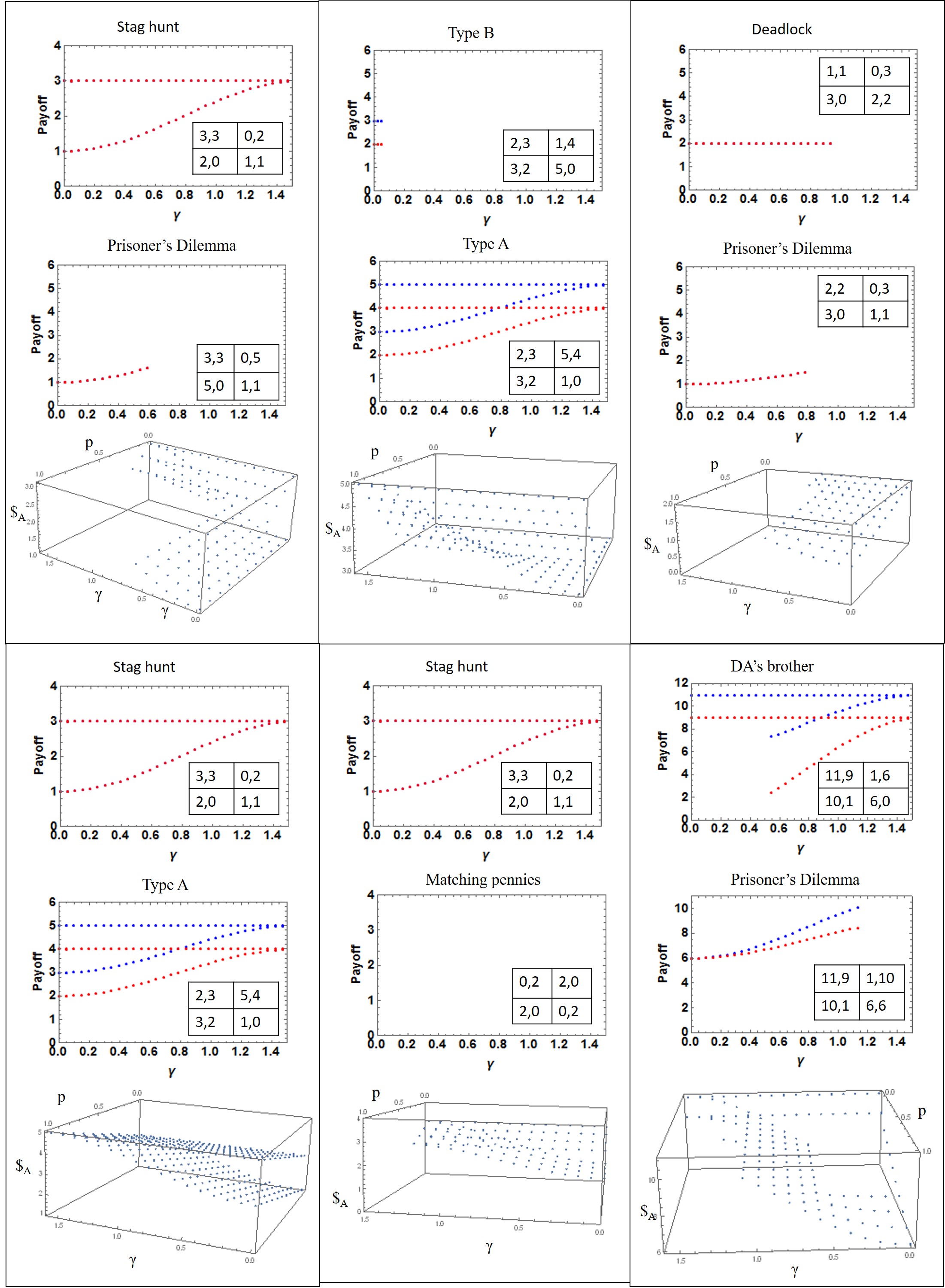}
\caption{\label{fig:BayesianData} Payoff at Nash equilibria for various games}
\end{figure}

The data are shown in Figure \ref{fig:BayesianData}, where a pair of two-player games are presented with their payoff matrices inset in the form of Table \ref{tab:PDmatrix}. In the case of asymmetric games, in the two-player games, player A is plotted in blue, and player B is plotted in red. The payoffs at the Nash equilibria are plotted for both players as a function of the entanglement parameter $\gamma$. 

The Nash equilibria curves come in two types, constant with $\gamma$, or increasing with $\gamma$. It is instructive to compare these to the solutions of the corresponding classical games. In all cases, the values of the payoff at no entanglement, i.e. $\gamma = 0$, match those of the classical game.  For games only have one constant Nash equilibria curve, such as in the games labeled `Type B,'  `Deadlock', and `Stag hunt,' there is one Nash equilibrium of the classical game and it is Pareto optimal. As the entanglement increases, the Nash equilibrium vanishes above some critical value.For games that have Nash equlibrium that grows with $\gamma$ (i.e. `Prisoner's dilemma') there is only one Nash equilibrium of the classical game but it is not Pareto efficient. The Nash equilibra for these games also vanish for some critical value of entanglement. The vanishing of the Nash equilibrium at a critical entanglement has been compared to a phase transition-like behavior\cite{Du2003}.

For games that have two Nash equilibrium in the classical game, one Pareto optimal, the Pareto optimal solution remains for all values of entanglement, whereas the second Nash equlibrium grows and converges with the optimal one at maximal entanglement.  In these cases, the Pareto optimal Nash equilibrium does not vanish at some critical entanglement. For two player games with no Nash equilibrium classically, such as the `matching pennies' game, there are no Nash equilibria in the corresponding quantum game.

In short, games with one equilibrium seem to lose that equilibrium at a critical entanglement, and in games with two equilibria, those equilibria persist for all values of entanglement.  Becuase our methods are numerical, these observations are not tantamount to formal proofs, and counter-examples may be found, but they are suggestive of a deeper structure. 
 
The `DA's brother' is an interesting outlier from this categorization. The classical game has only one Nash equilibrium, and it is Pareto optimal. However, as the entanglement increases, a second Nash equilibrium appears and then converges to the Pareto optimal solution as in the case of games with two equilibria mentioned above. Additionally, the Pareto efficient solution does not vanish at some critical value of entanglement.

\subsection{Bayesian game results}

Three-player Bayesian games can be composed out of a pair of two-player games. This can be interpreted as the player's having incomplete information about their opponents payoffs. The solutions to the Bayesian games composed of various two-player game combinations are plotted in 3D below the two-player results in Figure \ref{fig:BayesianData}. The payoff for only player A is plotted against the entanglement, $\gamma$, and the probability to play with each player, $p$ from Equation \ref{eq:Bpayoff}. The Bayesian graphs are oriented so that the top two-player game is in the back of the 3D plot with the bottom game in the front, and no entanglement on the right with maximal entanglement on the left.

As expected, the $p=0$ and $p=1$ solutions to the Bayesian game match the two-person game solutions for players A and B respectively. Along the $\gamma=0$ plane, the results match those of the classical game with mixed strategies. In games with the same number of Nash equilibria in the two component two-player games such as in `Deadlock' vs. `Prisoner's dilemma', the solutions at $p=0$ continuously and linearly transform into the solutions at $p=1$. 

When there are a different number of Nash equilibria in the two component two-player games, the equilibria must vanish, or appear, at some value of $p$. This is similar to the vanishing, or appearance, of Nash equilibria at a critical entanglement in the two-player games, only here, we also see them in the Bayesian game as the degree of incomplete information, $p$, changes.

The `DA's brother' vs. `Prisoner's dilemma' Bayesian game has been partially analyzed before\cite{Avishai2012}, and is the subject of a more detailed analysis that will be presented elsewhere\cite{QIP}.

\subsection{Towards continuous strategy choices}

\begin{figure}
\includegraphics[width=1\columnwidth]{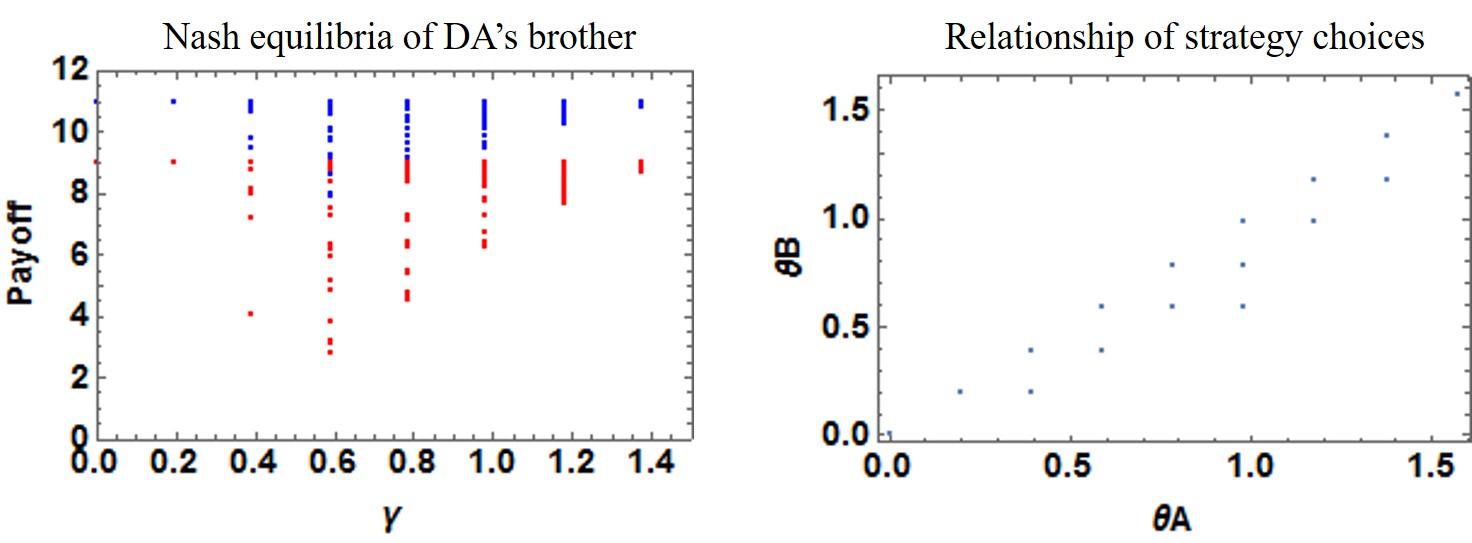}
\caption{\label{fig:continuous}DA's brother game with finer steps in strategy}
\end{figure}

Returning to a two-player game, we now examine the game as the discretization of the strategy choices in Equation \ref{eq:space} becomes finer, approaching the limit of completely arbitrary SU(2) rotations. The Nash equilibria of the `DA's brother' are now calculated using the stepping parameters $(\Delta \theta, \Delta \phi, \Delta \alpha)=(\pi/8, \pi/8, \pi/8)$ which yields 1824 unique strategy choices. As seen in the left graph of Figure \ref{fig:continuous}, the space between the two Nash equilibria becomes filled with many additional equilibria. The Nash equilibria found with $(\Delta \theta, \Delta \phi, \Delta \alpha)=(\pi, \pi/2, \pi/2)$ form the upper and lower bounds of the new Nash equilibria. In the right graph of Figure \ref{fig:continuous} the strategy parameters of the Nash equilibria for the players are compared by plotting $\theta_A$ vs $\theta_B$ against each other for each Nash equilibrium. This shows that the Nash equilibria generally follow the trend  $\theta_A = \theta_B$. 

\begin{figure}
\includegraphics[width=1\columnwidth]{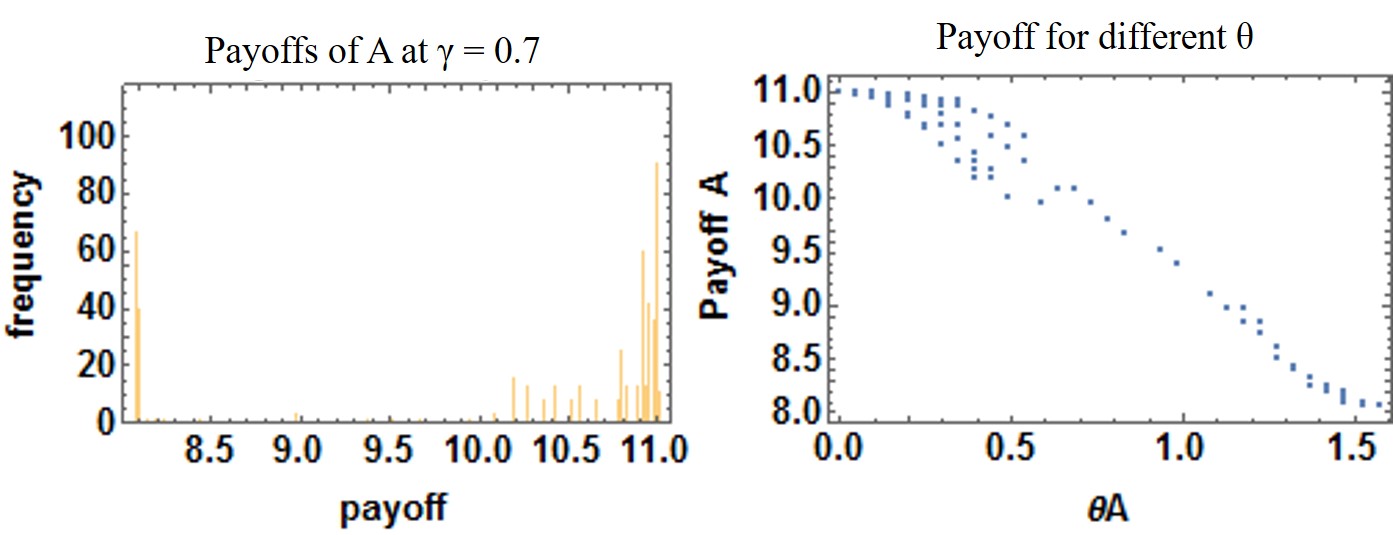}
\caption{\label{fig:continuous2} An analysis of the finer strategy steps at one value of entanglement.}
\end{figure}

Taking a slice of the payoff as a function of entanglement data at $\gamma =0.7$, a histogram of the payoffs achieved suggests that there is some structure within the distribution, as shown in the left hand graph of Figure \ref{fig:continuous2}. In this data the stepping parameters used were  $(\Delta \theta, \Delta \phi, \Delta \alpha)=(\pi/32, \pi/8, \pi/8)$, yielding 7968 unique strategy choices.  The data suggest that more Nash equilibria occur near the original Pareto optimal solution that occured with $\Delta \theta = \pi$. There may be some indication that the Nash equilibria are beginning to converge towards the two original Nash equilibria  However, as computation with finer steps is impractical with our current numerical method, further study is required. 

There is also a relationship between the payoff that is rewarded and the strategy choice at each Nash equilibrium. In the right side graph of Figure \ref{fig:continuous2} we plot the $\theta_A$ parameter of each Nash equilibria against the payoff of player A. When $\theta_A = \theta_B=0$, the payoff is the Pareto optimal solution, which is expected from the results with $\Delta \theta = \pi$. Then, as $\theta$ approaches $\theta = \pi$, the other equilibrium of the original game, the payoff transforms to the payoff of the second equilibria. Further study is needed to understand these relationships.

\subsection{Discussion and conclusions}

The Nash equilibria that arise in a quantum game, where entanglement produces correlations in the player's outputs, can be compared to the correlated equilibrium in classical game theory\cite{Auman1974}. A correlated equilibrium in classical games can arise when mixed strategies are used and there is communication between the players in the form of advice from a referee or a contract. If players receive some piece of advice, or react in a predetermined way to a random event, they can employ strategies that are correlated with one another and realize self-enforcing equilibria that are different from those in the mixed game without communication. 

When entanglement produces correlated outcomes for the players, the equilibria produced strongly resemble the correlated equilibria. In contrast to the classical case, the role of advice is played by the initial entanglement. And once that entanglement is imposed on the players by a referee, it forms an effective contract that cannot be broken. The correlations will persist even if the players are not allowed communication after the initial entanglement. In addition, quantum correlations can exhibit probability distributions that are not allowed by classical correlations, and can persist non-locally. 

The sudden dissapearance of a Nash equilibrium as the entanglement is increased suggests that the correlations can benefit the player's up to a point, but when the correlations are too strong, the Nash equilibrium no longer occurs. It would be interesting to find examples of classical games where the enforcement of some contract produces a benefit for the players, but if it is enforced too strongly, it ceases to allow a Nash equilibrium. In addition, the abrupt changing of the structure of Nash equilibrium as a function of the player's incomplete information could strongly effect any protocol on a network where the agents have some uncertainty about the payoffs or players in the game.  

% \disclosures 
\subsection*{Disclosures}
The authors have no relevant financial interests in the manuscript and no other potential conflicts of interest to disclose.

%\acknowledgments 

%%%%% References %%%%%
\bibliographystyle{abbrvnat}

%%%%% Biographies of authors %%%%%

\vspace{2ex}\noindent\textbf{Neal Solmeyer} is a physicist the Army Research Laboratory (Adelphi, MD). He received his  BA degrees in physic and philosophy from Carleton College (Northfield, MN) in 2006, and his PhD degree in physics from Penn State University (State College, PA) in 2013.  His research  interests include using Rydberg excitations in ensembles of laser cooled and trapped rubidium atoms for the purposes of quantum communication, and applying quantum game theory to distributed quantum networks. 
\vspace{1ex}
\noindent Biographies and photographs of the other authors are not available.

\listoffigures
\listoftables

\end{spacing}
\end{document}